\documentclass[aps,prb,titlepage,twocolumn,floatfix]{revtex4}

\usepackage{graphicx}
\usepackage{amsmath}
\usepackage[T1]{fontenc}
\usepackage[latin1]{inputenc}
\usepackage{times}
\usepackage[normalem]{ulem}
\usepackage{setspace}
  \makeatletter
  \def\@dotsep{4.5}
  \makeatother
\newlength{\myVSpace}
\newcommand{\ket}[1]{\left| #1 \right\rangle}
\newcommand{\bra}[1]{\left\langle #1 \right|}

\newcommand{\be}{\begin{equation}}
\newcommand{\ee}{\end{equation}}
\newcommand{\ba}{\begin{eqnarray}}
\newcommand{\ea}{\end{eqnarray}}
\begin{document}

\title{Mesoporous matrices for quantum computation
with improved response through redundance}
\author{T.E. Hodgson}
\affiliation{Department of Physics, University of York, Heslington, York,
YO10 5DD, United Kingdom}
\author{M.F. Bertino}
\affiliation{Department of Physics, University of Missouri-Rolla, Rolla, MO
65409, USA}
\author{N. Leventis}
\affiliation{Department of Chemistry, University of Missouri-Rolla, Rolla, MO
65409, USA}
\author{I. D'Amico}
\affiliation{Department of Physics, University of York, Heslington, York,
YO10 5DD, United Kingdom.}
%\pacs{03.67.Lx,73.21.La,81.07.Ta}

\begin{abstract}
We present a solid state implementation of quantum computation, which improves previously proposed optically driven schemes. Our proposal is based on vertical arrays of quantum dots embedded in a mesoporous material which can be fabricated with present technology. The redundant encoding typical of the chosen hardware protects the computation  against gate errors and the effects of measurement induced noise. The system parameters required for quantum computation applications are calculated for II-VI and III-V materials and found to be within the experimental range. The proposed  hardware may help minimize errors due to polydispersity of dot sizes, which is at present one of the main problems in relation to quantum dot-based quantum computation.
\end{abstract}

\maketitle
\section{Introduction}

The current high level of progress in the
design and manufacture of  low dimensional structures, has led to an increasing
interest in the development of solid state based quantum computing\cite{neilsenchang}
schemes/hardware.
Among the various proposals are schemes which rely on spin and exciton
qubits confined in semiconductor quantum dots (QD)\cite{PhysToday}, which can
be manipulated using ultra fast laser
pulses\cite{chapter}. Several of these
optical quantum computation
schemes rely on exciton-exciton direct Coulomb interaction, which provides the
necessary coupling to perform two qubit gates\cite{irene,gan}. The
presence of an exciton in a QD
produces a
biexcitonic shift in the ground state excitonic
energy of a nearby QD. By driving
a qubit at this shifted frequency conditional operations can be
performed\cite{irene,gan}.

In order to carry out practical quantum computation it is necessary to be able
to address individual qubits.
This poses a problem for driving optically the response of self assembled
quantum dot ensembles (such
as the ones grown by Stranski-Krastanow techniques): in these ensembles the size of each dot is one
order of magnitude smaller than the laser
spot addressing it.
To this end it  has been proposed to use energy selective
methods on isolated stacks of quantum dots (quantum registers)\cite{irene,gan}. However, it is still experimentally difficult to control QD  size and position in
a satisfactory way, and QD vertical stacks tend to form in the
plane at random positions.
Additionally the size of the QDs within the
stacks is hardly controllable, resulting in the practical difficulty of
creating the desired sequences of energy selectable excitonic transitions.

In the past years, techniques and materials have been developed that may allow to solve most of the fabrication issues associated with stacked quantum dot arrays. For example, materials like MCM-41 and SBA-15 consist of regular arrays of pores forming a hexagonal lattice\cite{889,696,8989}. By simple variations of the synthetic conditions, the pore diameter can be varied from a few nm to tens of nm. The thickness of the oxide walls separating the pores can also be varied, from about 1 nm to ca. 6 nm \cite{10001}. While the first materials of this kind were based on silicates, more recently matrices with a well-defined pore size and pore arrangement have been reported also for high dielectric constant materials such as ZrO$_2$ \cite{10002,10006}, and mixed Si-Ti oxides \cite{10003,10004,10005,nonhydro}.  Metal and semiconductor nanoparticles can be grown within the pores of these materials with techniques as varied as calcination \cite{676}, photolithography \cite{bertino,Gadipalli,Gadipalli2}, and electrochemistry\cite{1276,final}.  With these techniques, superlattices of quantum dots have been produced \cite{673}. The electrochemical route is probably the most interesting for the scheme that we propose. The group of M. Natan has demonstrated that stacked arrays of metals can be fabricated inside porous materials. Columns with a height of up to 15 $\mu$m made up by up to 8 stacked layers have been obtained \cite{1276}. In the future, it may be possible to fabricate composite materials made up of several layers of semiconductors disposed on a hexagonal lattice. The height of the dots within each layer will be controlled by the processing conditions (e.g., electrodeposition time) while the lateral size of the dots will be determined by the matrix pore size. The coupling between dots in different pores will be tuned by varying the wall thickness and/or by varying the dielectric constant of the oxide making up the walls.

\section{System and Redundant Encoding} \label{sectwo}
 We consider a system consisting
of a TiO$_2$
matrix in which alternating layers of two semiconductors with
widely different band gaps are deposited. This
provides a stack of QDs (qubits), sandwiched between the larger band
gap material (barriers).
The resulting  system is depicted in Fig.~1(a) and
consists of an array of identical, hexagonally packed
stacks of quantum dots (columns).
The band structure within each column is sketched in Fig.~1(b). As shown  in sec.~\ref{secthree}
the distance between nearby columns, and the high dielectric constant ($\epsilon \approx 100$)\cite{TiO2} of the matrix  is sufficient to consider each stack as
isolated from its neighbors.

\begin{figure}[h]
\includegraphics*[scale=0.4]{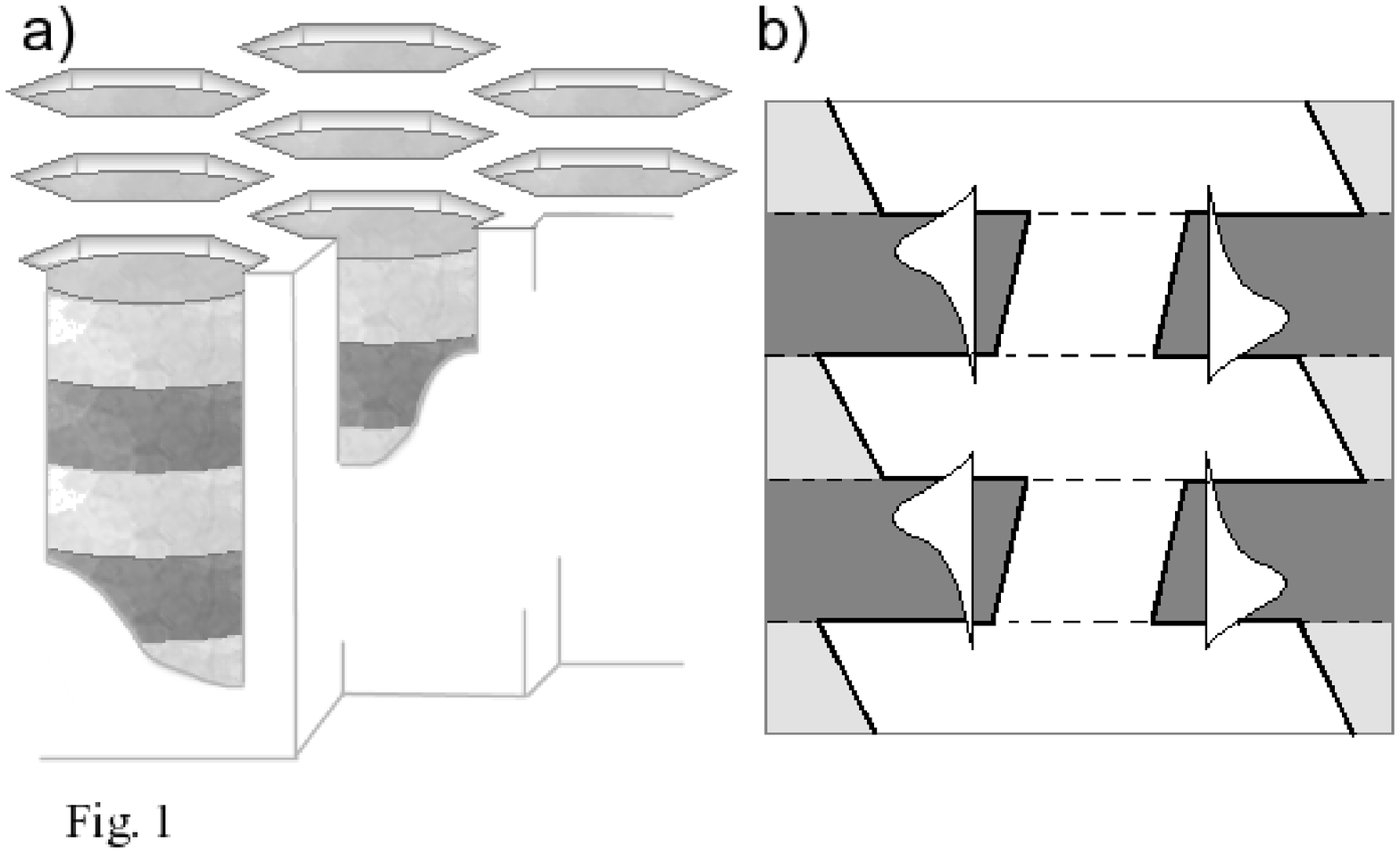}
\caption{Fig.~1: (a) The proposed system, an ensemble of stacks of alternating QDs (dark shade) and barriers (light shade) and (b) the band structure of each individual stack including the intrinsic field of the materials.}
\label{fig:structure}
\end{figure}

We propose to use semiconductors which can assume
wurtzite crystal structure. This leads to a strong intrinsic
electric field\cite{piezo}, which enhances coupling between
excitons in neighboring quantum dots {\it within a stack}\cite{gan}. 
We calculate the  built-in electric fields inside the quantum dots by considering an alternating sequence of quantum wells and barriers: results are in good agreement with experimental findings\cite{gan}, showing that the lateral shape of the dot is mainly responsible for the strong in-plane carrier confinement.
Quantum computation can then be carried out by using sequences of
laser pulses, as described in Ref.~\onlinecite{chapter}. Under the influence
of the same laser
pulse, each column will act as an independent replica of the same computational
array. The advantage of the hardware we propose is this intrinsic
redundance.

A practical quantum computing scheme must include some error-correction
strategy
for errors due to computation or hardware faults. A possibility is to average
over
many individual occurrences of the same quantum algorithm, so that
fluctuations around the expected result are protected against. An
example of this is seen in nuclear magnetic resonance (NMR) schemes\cite{NMR},
where large
ensemble of qubit arrays are naturally available.  In the current work
we propose a somewhat analogous, but solid state based, ensemble.
The ensemble is constituted by the quantum dot columns uniformly distributed
within the matrix. The advantage of our solid state ensemble over
NMR ones is twofold, namely the ability to initialize  all arrays in the
ensemble to a known well defined state -- the 'no-exciton' ground state -- and
the  intrinsic order of our ensemble,  which e.g. allows for a certain
degree of {\it spatial} addressability.
In this respect we mention that on the micrometer scale, different
semiconductors can be deposited with photolithographic techniques
on areas of the matrix, to create
regular supra-arrays of selected geometries, e.g. hexagons or stripes\cite{bertino,Gadipalli,Gadipalli2}. We foresee that this property might be used to perform different calculations
on different areas of the matrix.
The envisaged possibility of growing relatively long arrays, i.e. quantum registers of the
order of some tens of qubits (QDs), is another advantage over NMR based systems.

\section{Theoretical Model and Calculated Parameter Range} \label{secthree}
To check the feasibility of the proposed scheme, we model each individual stack
as a column of cylindrical quantum dots with
the same radius. To calculate the biexcitonic shifts, the confining potentials
are modeled as parabolic potentials with the same
characteristic widths as the expectation values $\sqrt{<z^2>}$ and $\sqrt{<r^2>}$ of the cylindrical dots in the stack.  Nearby quantum dots are coupled by the
biexcitonic shift $\Delta E $ between ground state excitons. In order to calculate the correct parameter space, we approximate the
biexcitonic shift $\Delta E$ as
\begin{equation}
| \Delta E|=|\bra{\psi_1\psi_2}U_C\ket{\psi_1\psi_2}|,\label{integral}
\end{equation}
where $\psi_i(r_{ie},r_{ih})=\psi_{ie}(r_{ie})\psi_{ih}(r_{ih})$ is the
wavefunction of the exciton in QD$_i$ in the single particle approximation, and
$U_C$ is the Coulomb interaction between the two excitons.

In the proposed hardware, the stacks are separated by the matrix walls, which can currently be made up to $6$~nm thick whilst maintaining the order of the structure. To be negligible, the inter-stack interaction must be much smaller than the interaction between excitons within a column,  i.e. $|\Delta E_{IC}^{tot}|<<|\Delta E|$. 
As we will show below, both II-VI and III-V based systems can be designed so that there is more than  an order of magnitude difference between the two energy scales, whilst
mantaining the matrix walls within the experimental range. 

 If all the stacks in the ensemble were to compute correctly -- i.e. no computational errors -- inter-stack interactions would only renormalize the  excitonic energies by the same amount for each stack, which could be easily accounted for in the computational scheme. However the event of computational failure in a certain stack will induce  a local, unwanted, shift of the exciton energies in neighbouring stacks. Each individual stack will interact with a certain number of such 'faulty computing' stacks. Therefore the magnitude of this unwanted shift is the sum of the interaction energies of a resident exciton with an exciton in each 'faulty computing' stack.  
 The stacks are arranged in a hexagonal structure, which can be represented as a series of concentric hexagonal shells surrounding each stack. The $i$~th shell consists of $6i$ stacks. The number of expected computational failures in the $i$~th shell will therefore be $6i\cdot(1-p)$, where $p$ is the probability of a successful computation for any individual stack. We estimate the total energy shift due to interactions with all the 'faulty computing' stacks in the ensemble as:
\begin{equation}
|\Delta E_{IC}^{tot}|=\sum_{i=1}^{ensemble} 6i(1-p)\Delta E_{IC}(r_i) \label{totalinter}
\end{equation}
where $r_i=(r_{max}+r_{min})/2$ is radius of the $i$~th hexagonal shell, $r_{max}$ and $r_{min}$ being the maximum and minimum distance of the shell from the central stack. $\Delta E_{IC}$ is calculated according to Eq.~(\ref{integral}).

\begin{figure}[h]
\includegraphics*[scale=0.9]{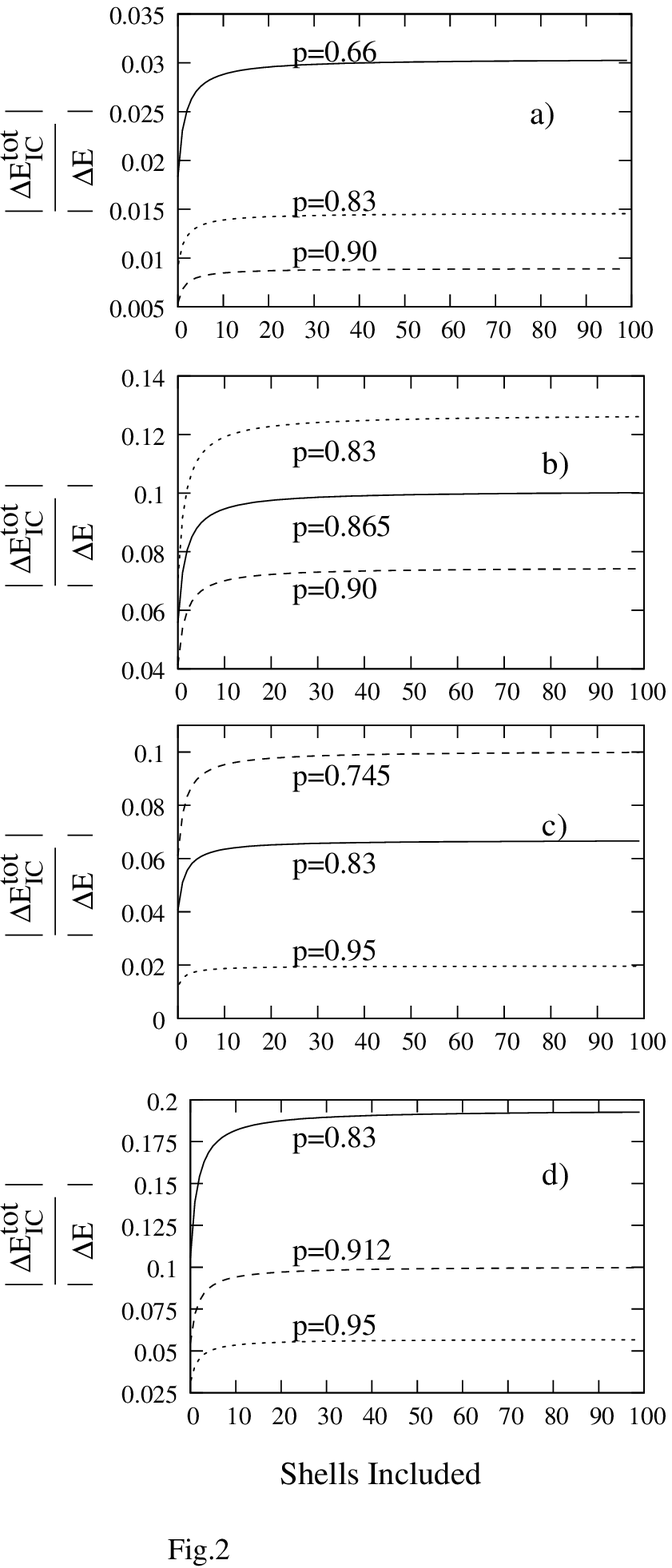}
\caption{Fig.~2: Ratio of {\it inter}-column biexciton interaction energy, to {\it intra}-column biexciton interaction energy ($\frac{|\Delta E_{IC}^{tot}|}{|\Delta E|}$) for different probabilities of success $p$ for any individual stack, against the number of shells of neighbors included. a) GaN/AlN  system with $6$~nm walls, b) CdSe/CdS system with $6$~nm wall, c) GaN/AlN system with 2nm walls, d) CdSe/CdS system with $4.5$~nm walls.}
\label{fig:inter}
\end{figure}

In Fig.~2 we show the results for system parameters appropriate for implementing computational schemes. As a typical GaN/AlN system,  we consider
$5$~nm porous radius with QD height $3.2$~nm and barrier width $2.4$~nm; As a typical CdSe/CdS system we consider $5$~nm porous radius with QD height $10.75$~nm and barrier width $5$~nm. It will be shown later in this section that these parameters are indeed  appropiates to our scopes.  Fig.~2(a) and Fig.~2(b) correspond to the GaN/AlN and CdSe/CdS systems respectively with matrix wall thickness of $6$~nm. It can be seen that for GaN QDs (panel a), each  stack in the ensemble can be considered isolated for $p=0.66$ and indeed for whichever $p$ value (not shown). 
$p=0.66$ corresponds to an average of two failures among  the nearest neighbors.
For CdSe QDs (panel b)  the ensemble can be considered isolated for $p\stackrel{>}{\sim}0.86$. Fig.~2(c) and Fig.~2(d) show the same calculation but for different wall thicknesses.  
In the GaN/AlN system even for matrix walls as thin as $2$~nm and $p\stackrel{>}{\sim} 0.74$, stacks can be still considered isolated  (panel c); for the CdSe/CdS system with matrix walls of $4.5$~nm (panel d), the individual stacks can be considered isolated for $p\stackrel{>}{\sim}0.91$ only.  Our results show that that each system can be designed so that individual stacks can be considered isolated for realistic wall thicknesses and 
a probability of success for a single stack which is reasonably low in CdSe/CdS and can be {\it arbitrarily low} in GaN/AlN.

The biexcitonic shift between neighbouring qubits within a stack can be
exploited to perform
two-qubit gates using multicolor train of laser pulses\cite{irene,gan,paulibl}.
For performing
operations on picosecond time scales -- which is essential due to the
relatively short excitonic decoherence times --  $\Delta E $  must be
of the order of a few meV. When choosing the correct parameter range however,
additional factors must be taken into consideration\cite{irene}.
 Larger $\Delta E $ can be induced by increasing the height of
the quantum dots. This increases the excitonic dipole moments
 under the influence of the intrinsic electric field. Care must be taken
however in
allowing at the same time for a satisfactory oscillator strength. Finally the
barrier width must be
large enough to ensure that single particle tunneling  between stacked
quantum
dots is negligible on the relevant time scales. The tunnelling time is calculated by taking the inverse of the tunneling rate in Ref.~\onlinecite{storage}.

 Fig.~3 shows the range of barrier widths and QD (qubit)  heights which satisfy  all
of the above conditions. The constraints on the system paramaters are $\Delta E > 3meV$, oscillator strength $\mu>0.15\mu_{zerofield}$, tunneling time $\tau>1ns$ and $(\frac{|\Delta E_{IC}^{tot}|}{|\Delta E|})<0.1$, where the latter is the ratio between the {\it inter}-column to the {\it intra}-column exciton-exciton interaction energy.

\begin{figure}[h]
\includegraphics*[scale=1]{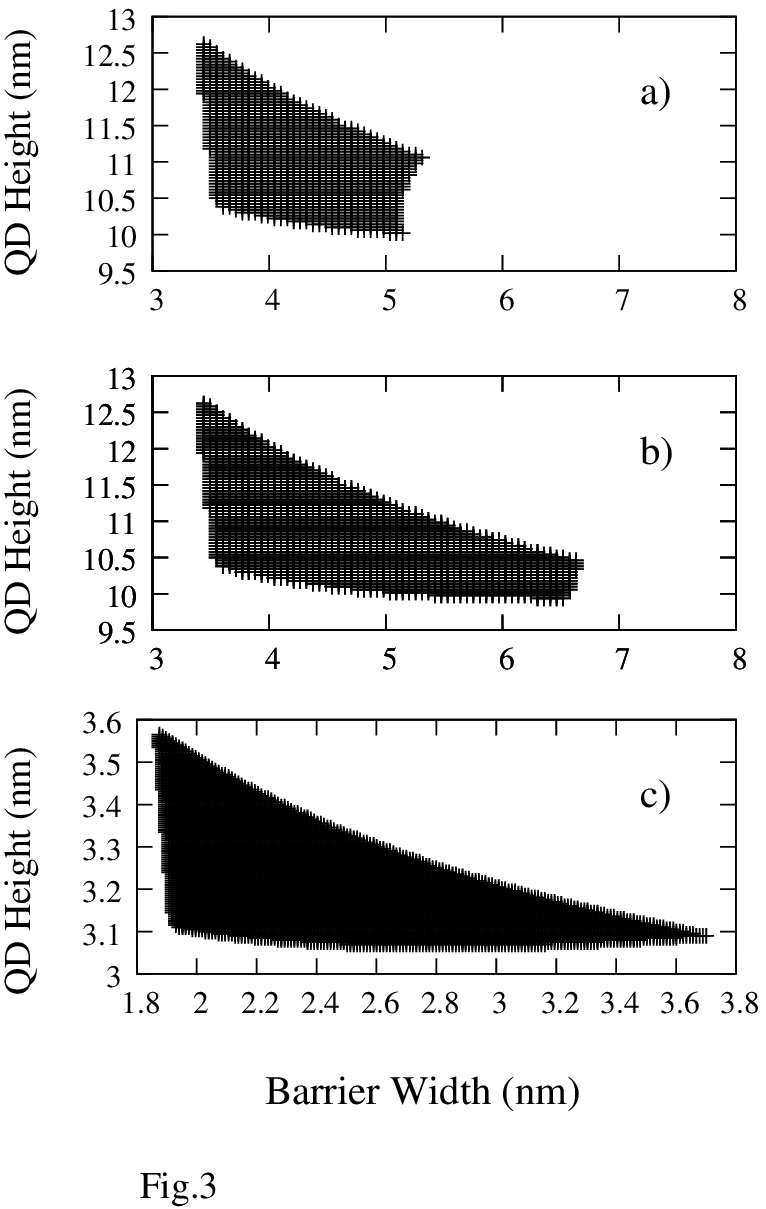}
\caption{Fig.~3: Parameter space of QD height and barrier width for which $\Delta E > 3meV$, oscillator strength $\mu>0.15\mu_{zerofield}$, tunneling time $\tau>1ns$, and $(\frac{|\Delta E_{IC}^{tot}|}{|\Delta E|})<0.1$  for CdSe/CdS (panels (a) and (b)) and GaN/AlN (panel (c) stacks of radius 5nm.}
\label{fig:parameterspace}
\end{figure}

 We have considered both II-VI and III-V
heterostructures, specifically
CdSe QD and  CdS
barrier (panels a and b), and GaN QD and AlN barrier (panel c). Panel a) shows the parameter space for a CdSe/CdS system with $4.5$~nm matrix walls, and $p=0.91$. Panel b) corresponds to the same system but for $p=0.93$. It can be seen that the parameter space is reduced by the requirement of negligible interaction among different QD stacks. Decreasing $p$ and/or the wall thickness increases the inter-column interaction, so that the region in parameter space corresponding to low values of $\Delta E$ is 'cut off'. For large $p$ and/or wall thickness the shape of the CdSe/CdS space would be similar to the GaN/AlN case (panel c). For a wall thickness of $4.5$~nm this occurrs at $p\approx 95$.  For reasonable values of $p$ and  wall thickness, due to the low interstack coupling in the III-V system, the request that  different stacks do not interact does not affect the parameter space.
Our results show  that for both
systems there is a wide range of dot sizes, which produce a suitably
large biexcitonic shift in the absorption spectrum of the quantum dot. The
compatible parameter space in the II-VI system corresponds to taller QDs than the III-V case. This is due to the intrinsic
electric field being smaller in the CdSe/CdS system  than in  GaN/AlN.
Therefore larger dot heights and barrier widths are needed to displace the electron and hole wavepackets to provide sufficient coupling between excitons.

Synthesis of II-VI QDs is generally speaking easier than synthesis of of their III-V counterparts. However it has been shown\cite{efros} that due to the
electron-hole exchange interaction the
ground excitonic state of materials with wurtzite structure is  optically
passive, the separation between bright and dark exciton being too large in
II-VI systems to be
negligible in respect to the energy scales we consider.
To overcome this problem, we suggest to consider an n-doped structure
such that each quantum dot traps a single electron. As
demonstrated experimentally for III-V materials\cite{experimentalndoped}, the
exchange splitting can in this way be switched off.
This would allow for implementation of schemes such as the one described in
Ref.~\onlinecite{paulibl}, where
the spin of the excess electron is the qubit and excitons are used for two
qubit gating.  It has been shown that  for dots in the strong coupling regime, the wave function of the ground state exciton is not significantly affected by the presence of an extra electron\cite{paulibl}.  Its effect on the biexcitonic shift, which depends on the shape of the excitonic wave functions, can then be safely neglected. 
Experimentally a possible way of doping each dot with exactly a single electron has been found by doping 
a QD ensamble with an electron  density which matches the dot density. Due to the strong Coulomb repulsion, double occupancy of the dot is avoided. A similar solution could be used to dope the dots in our structure. We envisage an alternative method, which could improve also detrimental stochastical effects.  We suggest to apply a bias between the top and the bottom of the structure. In this way each dot would be occupied by exactly one dot starting from the bottom and upwards. For not too strong biases Coulomb blockade would in fact prevent more than one electron occupaying each dot. Once the process has been completed,  the bias would be removed and the computational process could begin. As a possible alternative to doping, we can consider
CdSe QDs with radii greater than 5nm, for which
the exchange splitting becomes very  small\cite{efros}. These QDs fit well into
the parameter
space described in Fig.~3.

We underline that both  II-VI and III-V semiconductor structures can be grown
in mesoporous matrices with
present technologies\cite{10000,final,676}.
\section{Protection against errors and noise through redundant encoding} \label{secfour}
As shown above, each individual stack in the structure we propose
 will behave as an isolated computational register in which quantum
operations such as entanglement can be carried out using
 an appropriate sequence of (sub)picosecond laser
pulses\cite{irene,gan,paulibl}. An appropriate modulation of the QD
heights (done as the stack is grown) will result in a different ground
state exciton energy (or sequence of energies\cite{Lovett}) for each QD in a stack. This allows each qubit in a stack to be selectively addressed and
arbitrarily rotated around the Bloch sphere by laser pulses of the appropriate
frequency, duration and phase. Similarly conditional two-qubit operations (e.g.
entanglement)
can be performed\cite{irene,gan,paulibl}.

Our hardware could be used to implement quantum algorithms.
The final phase of the algorithm would be to make a
measurement on the qubits, generally in the computational basis, to obtain the
'answer'. The measurement of an $n$ qubit output, consists of $n$ individual
measurements, each of which will be found in either the $\ket{1}$ or $\ket{0}$
state. For each individual qubit, being part of an ensemble provides protection against
computational errors and measurement noise.  In the following discussion we will focus for simplicity, on a single qubit output and assume that before measurement, the qubit is stored in a
particular QD which we will refer to as  a 'storage' qubit\cite{storage}.
For simplicity we will think of it as the upper QD in the stack, though this is {\it not}
a necessary condition.

After initial preparation (all qubits are initialized in the $\ket{0}$
state, i.e. no excitons present),
the driving laser beam will illuminate a circular section of hexagonally
packed stacks: this ensemble of $N$ stacks
 represents our redundantly encoded ensemble, since the train of laser pulses
will simultaneously
drive the same  operations on all the stacks of the ensemble. Finally the
result from
the ensemble is stored in the  $N$ upper (storage) qubits and may be read off.

 In the event of a perfectly successful algorithm with no errors, the entire
ensemble of storage qubits would all be in the correct state, which we will
assume without loss of generality, to be the $\ket{1}$ state. In reality
however, there is a possibility that the quantum algorithm will fail on any one
stack. This leads to $n\le N$ storage
qubits being in the correct state after the computation. Let us assume that in
measuring  the storage qubit ensemble, the output signal (e.g. photons from
excitonic recombination, variation of current through a narrow contact...) is
proportional to $n/N$ and in particular a signal $I$ will be measured
within the range $I_{min}$ (corresponding to all storage bits being in the
'wrong' state, e.g. the
$\ket{0}$ state) to $I_{min}+\Delta I$ (corresponding to the whole ensemble in
the
$\ket{1}$ state). In the hypothesis that the quantum registers (columns) are
uncorrelated, the actual signal would be  then
$I= \Delta I  n/N$ where we have set the zero signal at $I_{min}$. There
will also be fluctuations about this value due to
noise
induced during measurement. In the following we derive a simple relationship to estimate the
required ensemble size to correct for given  error probabilities. In
particular we want to
demonstrate that with a modest ensemble of $N\approx100$ (corresponding to the
area illuminated by a laser at optical frequencies, with a spot  of diameter $\sim 10^3 $\AA), 
it is possible to correct for
sizable errors both in the computation and due to measurement noise.

Let us assume that the probability $p$ of a successful computation in an
individual stack of
qubits is constant across the entire ensemble. For an ensemble of $N$ stacks of
qubits, the probability distribution of obtaining $n$ correct answers is
binomial
\begin{equation}
P(n)=\left( \begin{array}{c}
N\\n
\end{array} \right) p^n(1-p)^{N-n} \label{eq:bin}
\end{equation}

If we assume that in the event of a failure the state of the qubit is
 found to be in the $\ket{0}$ state with
probability $q$, the probability distribution becomes

\begin{eqnarray}
P(n)&=&\sum_{i=0}^{n}\left( \begin{array}{c}
N\\n-i
\end{array} \right) p^{n-i}(1-p)^{N-(n-i)} \cdot \nonumber \\ & &\left(
\begin{array}{c}
N-(n-i)\\i
\end{array} \right) (1-q)^i q^{N-n}
\end{eqnarray}
where $i$ denotes how many failed computations have been randomly assigned the
correct output. This can be shown to be equal to
\begin{equation}
P(n)=\left( \begin{array}{c}
N\\n
\end{array} \right) [p+(1-q)(1-p)]^n[q(1-p)]^{N-n}, \label{eq:distribution}
\end{equation}
which is of the same form of Eq.~(\ref{eq:bin}).
The shift of the mean value and standard deviation, from those of the
probability distribution (\ref{eq:bin}) is due to the fact that for a binomial
distribution, even in
 in the event of a failure there is a finite probability $(1-q)$ of obtaining
the correct result.
For a large ensemble of dot stacks, the binomial distribution can be
approximated by a Gaussian distribution.
The probability distribution~(\ref{eq:distribution}) under the Gaussian
approximation yields a mean value of $\bar{n}=N[1-q(1-p)]$ and a standard
deviation of $\sigma=\sqrt{N[1-q(1-p)][q(1-p)]}$. If the size of the
ensemble $N$ satisfies
\begin{equation}{\frac{N}{2}}<\bar{n}(N)-k\sigma(N),\label{treshold1}\end{equation} then the result
$I>\Delta I/2$ (i.e. $n>N/2$) from a
{\it single measure} of the ensemble
will indicate that the answer from the computation is  $\ket{1}$  ($\ket{0}$
otherwise) with probability $99.7$\% for $k=3$.

\begin{figure}[t]
\includegraphics*[scale=1]{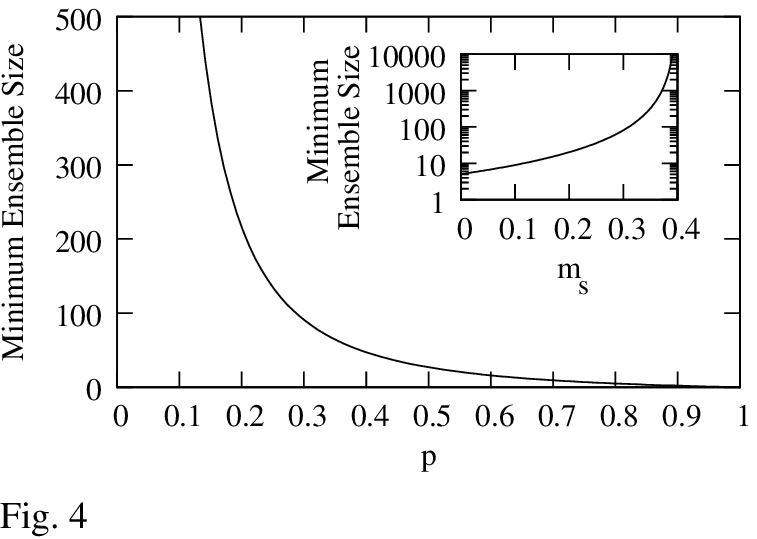}
\caption{Fig.~4: Minimum ensemble size vs probability $p$ with $q=1/2$ and $m_s=0$. Inset: Minimum ensemble size vs noise $m_s$ for $p=0.8$ and $q=1/2$.}
\label{fig:noise}
\end{figure}

In Fig.~4 we plot the minimum value of N which satisfies Eq.~(\ref{treshold1}) for $k=3$ in respect to the probability of individual success
$p$.
Here we assume  no
systematic bias, so we set $q=1/2$.
 The figure shows that, {\it due to the redundant
encoding},  even when each individual stack
computes correctly with a probability as low as $p=0.3$, an ensemble of
$N\stackrel{<}{\sim} 100$ stacks
is sufficient for measuring the
correct answer with such a high confidence. As discussed in section III, the values of $p$ for which our computational scheme applies   depend on the thickness of the matrix walls and system materials. For a CdSe/CdS system with $4.5$~nm matrix walls the computational scheme only works for $p>0.912$, however for a GaN/AlN system with $6$~nm walls the scheme is valid for all $p$.

Let us now consider that experimentally there will always be a certain  amount
of noise associated to the
measurement,  and discuss how the redundant encoding can help in tolerating
this source of error.
We can describe this noise by modifying
Eq.~\ref{treshold1} as
\begin{equation}
N\left({\frac{1}{2}}+m_s\right) < \bar{n}(N)-k\sigma(N)\label{eq:relation}
\end{equation}
where $m_s={\Delta I_{noise}/\Delta I}$ and $\Delta I_{noise}\equiv
2\mbox{max}\{|I-\Delta I (n/N)|\}$.
Again, if N satisfies Eq.~(\ref{eq:relation}) with $k=3$, then a measured
signal  $I>\Delta I/2$ ensures that the
result of the algorithm is $\ket{1}$  with a probability of $99.7$\%. By
rearranging
Eq.~(\ref{eq:relation}) we obtain  the condition

\ba
& & N>\frac{k^2[1-q(1-p)]q(1-p)}{\{{\frac{1}{2}}+m_s-[1-q(1-p)]\}^2}\label{eq:relation2} \\
& &\mbox{with}~\left({\frac{1}{2}}-m_s\right)>q(1-p) .\label{eq:relation2:cond}
\ea
Setting $m_s=0$ in Eqs.~(\ref{eq:relation2}) and (\ref{eq:relation2:cond}),
gives
 a lower bound on the ensemble size for the case of a noiseless
measurement. For $m_s\to 1/2- q(1-p)$  the size of
the ensemble needed to correct for this noise
tends to infinity.
The inset of figure~4 shows that  for an ensemble of N=100
stacks, for $p=0.8$ and $q=1/2$,
the system is robust even for a measurement noise as high as $30\%$.

We underline that, using the proposed fabrication method, $N\approx 100$
corresponds to the
smallest laser beam spot, solving the problem of spatial addressability in
 QD-based quantum computing schemes.
\section{Conclusions}\label{secfive}
A scheme for implementing quantum algorithms with improved
response through redundancy has been presented. Our proposal is based on a
mesoporous
matrix which provides an uncorrelated
 ensemble of computational arrays. Our method protects
against both computational errors
and measurement induced noise, and,  using relatively
small ensemble sizes, correct answers are found with
probability greater than 0.997.
Our hardware and computational scheme alleviate many issues of quantum computing schemes based on semiconductor QDs.
In these schemes, a major error source is the uncontrolled polydispersity of
dot sizes due to the experimental growing techniques. This directly affects the
quantized energy levels, detuning them from the ideal ones. We foresee that our
scheme can provide a feasible way for containing this source of error.

\end{document}